\documentclass[]{IEEEphot}
\usepackage[normalem]{ulem}

\begin{document}

\title{High-speed Implementation of Length-compatible\\ Privacy Amplification in Continuous-variable Quantum Key Distribution}

\author{Xiangyu Wang$^1$, Yichen Zhang$^1$, Song Yu$^1$, and Hong Guo$^{1,2}$}

\affil{$^1$State Key Laboratory of Information Photonics and Optical Communications, Beijing University
of Posts and Telecommunications, Beijing 100876, China\\
$^2$State Key Laboratory of Advanced Optical Communication Systems and Networks, School of Electronics
Engineering and Computer Science, Center for Quantum Information Technology, Center for Computational
Science and Engineering, Peking University, Beijing 100871, China}

\maketitle

\begin{receivedinfo}%
This research was sponsored in part by the Key Program of National Natural Science Foundation of China under Grant 61531003, in part by the National Natural Science Foundation under Grant 61427813, and in part by the National Basic Research Program of China (973 Pro-gram) under Grant 2014CB340102. Corresponding author: Song Yu (e-mail: yusong@bupt.edu.cn); Yichen Zhang (e-mail: zhangyc@bupt.edu.cn).
\end{receivedinfo}

\begin{abstract}
Privacy amplification is an indispensable step in postprocessing of continuous-variable quantum key distribution (CV-QKD), which is used to distill unconditional secure keys from identical corrected keys between two distant legal parties. The processing speed of privacy amplification has a significant effect on the secret key rate of CV-QKD system. We report the high-speed parallel implementation of length-compatible privacy amplification algorithm based on graphic processing unit. Length-compatible algorithm is used to satisfy the security requirement of privacy amplification at different transmission distances when considering finite-size effect. We achieve the speed of privacy amplification over 1 Gbps at arbitrary input length and the speed is one to two orders of magnitude faster than previous demonstrations, which supports high-speed real-time CV-QKD system and ensures the security of privacy amplification.
\end{abstract}


\section{Introduction}

Quantum key distribution (QKD)~\cite{QKD} is one of the most practical quantum information technologies, which allows two distant legitimate parties named Alice and Bob to generate unconditional security keys despite the presence of an eavesdropper Eve. According to the encode mode of key information, there are mainly two kinds of QKD systems: the systems that encode key information on discrete-variable (DV)~\cite{BB84,E91} and the systems that encode key information on continuous-variable (CV)~\cite{GG02,2004NOSW,2014LZ,2015SP,2014ZYC}. We focus on CV-QKD system because it has the advantage of using standard telecommunication technologies~\cite{2012RMP,Paul13,2015en,Zhang2017}. A practical CV-QKD system includes two parts: quantum communication and classical communication. In the first part, the sender Alice prepares quantum states and sends them to receiver Bob though an non-trusted quantum channel which can be eavesdropped by Eve, then Bob measures the quantum states by homodyne or heterodyne detector. Generally, the second part is the postprocessing of CV-QKD systems, which mainly contains sifting, parameter estimation~\cite{PE1,PE2}, information reconciliation~\cite{IR1,WangRA,WangHSEC} and privacy amplification~\cite{PA1,PA2,PA3}, and it is implemented though a classical authenticated channel.

Privacy amplification is an indispensable step in postprocessing, which is used to distill the final secret keys from the identical corrected keys between Alice and Bob. This process is implemented by using universal Hash families which can be used to compress keys~\cite{Hash1,Hash2}. Toeplitz matrix is one of universal hash families~\cite{Toeplitz}. Its structure is simple and can be implemented in parallel. For high speed real-time CV-QKD system, the implementation speed of privacy amplification is one of the limitations. Some approaches can be used to speed up the process based on software implementation, such as fast Fourier transform (FFT) and number theoretic transform~\cite{NTT}. There are some other methods have been used to accelerate the process. In Ref.~\cite{ZhangCM}, they use four basic multiplication algorithms at different input length to achieve a fast implementation. In addition to software implementation, this process can also be implemented with hardware, such as field-programmable gate arrays (FPGA)~\cite{FPGA}.

Currently, the main bottleneck of CV-QKD system is the postprocessing of information, especially in high efficiency and high speed real-time security implementation, and it highly affects the secret key rate and transmission distance of CV-QKD systems~\cite{long11,Slice1}. We achieve high efficiency by combining multidimensional reconciliation~\cite{MR} and multi-edge type low density parity check codes (MET-LDPC)~\cite{METLDPC}. However, the MET-LDPC codes require long block length (on the order of $10^6$), large iteration numbers and soft-decision decoding algorithm to correct errors at low signal-to-noise ratios~\cite{WangRA}. Generally, it is more suitable for software implementation which has the advantages of big memory and high floating-point precision compare to hardware-based decoder. We obtain high error correction speed for CV-QKD system based on graphic processing unit (GPU), because it provides parallel computation and is suitable for processing floating-point information~\cite{WangHSEC}. To obtain high speed real-time CV-QKD systems and simplify the postprocessing operations, we still implement high speed privacy amplification on GPU. Due to the high computational complexity of matrix multiplication, we convert the matrix multiplication into polynomial multiplication and accelerate it with FFT which reduces the computational complexity from O($n^2$) to O($nlog_{2}n$).

Furthermore, the input length of privacy amplification is required to be large enough to ensure the security of CV-QKD systems when considering the finite-size effect~\cite{PE1,ZhangXY}. Normally, the input length increases as the transmission distance increases when the security parameter is fixed. The security parameter represents the failure probability of the system, which means that the eavesdropper Eve can get the secret keys with a probability no large than the value of the security parameter. Typically, it is fixed to $10^{-10}$. Practically, when the transmission distance are about 50km, 80km, and 100km, the input length are respectively the order of $10^8$, $10^9$, and $10^{10}$~\cite{Paul13,Zhang2017}. However, the computational power of common computers is limited by memory size or other factors, and can not perform long-input-block-length privacy amplification directly. Thus, we propose a length-compatible method to satisfy the length requirement of different transmission distance. This method is implemented by dividing the long input length into short block length, and then performing the privacy amplification separately, finally merging the corresponding results together. The correctness of FFT results is related to the input length and computational precision. With the increase of input length, the required computational precision of FFT increases as well. Otherwise, the results will be wrong and the privacy amplification will fail. And the higher the precision, the slower the speed. To solve this problem, we separate the long input length into short block length which can be correctly calculated at low precision. Meanwhile, in order to make full use of the computing and memory resources of GPU and further accelerate the speed of privacy amplification, the short blocks can be calculated in parallel by setting batches.

In this paper, we propose a high speed implementation method of length-compatible privacy amplification for CV-QKD system. The early stage of this work has been applied to the longest field test of a high speed real time CV-QKD system and we further improve it in this work~\cite{Zhang2017}. The principe of privacy amplification for CV-QKD system is described in Section 2. In Section 3, we present the high speed implementation method of length-compatible privacy amplification based on GPU. The results under different input length are given in Section 4. And this paper is concluded in Section 5.

\section{The Principle of Privacy Amplification for CV-QKD system}

Privacy amplification is one of the important steps in symmetric-key quantum cryptography by providing an approach for Alice and Bob to distill unconditional security keys from a shared corrected keys in order to securely encrypt/decrypt information. This step is realized over a classical public channel, but it is assumed to be authenticated, so that adversary Eve can eavesdrop on the messages exchanged between legitimate parties Alice and Bob but she can not modify them without being discovered. Typically, privacy amplification algorithm is achieved by universal hash families which provides information-theoretically secure for CV-QKD system. For a practical system, the finite-size effect on the security of privacy amplification can not be ignored and the size is related to the transmission distance, security parameter, and other factors. In this section, we first introduce the universal hash family used in our system, then describe the finite-size effect of privacy amplification, finally present the implementation procedure of privacy amplification algorithm.

\subsection{Universal Hash Families}

An important application of universal hashing is privacy amplification in quantum cryptography. A hash function is selected at random from universal hash families to extract secure keys from corrected keys at a low collision probability. The collision probability means that different input data have the same output keys after universal hashing. In other words, even if adversary Eve does not have the same corrected keys as Alice and Bob, she can also have the same secret keys with a certain probability. Thus, collision probability should be as small as possible, otherwise the security of the secret keys can not be guaranteed. And its value is determined by three factors: the hash function that selected from universal families, the input length and the output length of the hash function. Different hash functions have different implementation complexity. Thus, choosing a hash function with low collision probability and low complexity is of great significance to the security and the speed of privacy amplification for CV-QKD system.

Toeplitz matrix is one of the universal hash functions, it is also called diagonal-constant matrix which means that the element on a diagonal line from the upper left to the lower right of the matrix is constant. Therefore, all elements of the whole matrix are determined as long as the elements of the first column and the first row of the matrix are determined. For instance, Eq. (\ref{toeplitz}) is a Toeplitz matrix.

\begin{equation}
T={
\left[ \begin{array}{cccccc}
t_0 & t_n & t_{n+1} & \cdots & \cdots & t_{2n-2}\\
t_1 & t_0 & t_n & \ddots & & \vdots\\
t_2 & t_1 & \ddots & \ddots & \ddots & \vdots\\
\vdots & \ddots & \ddots & \ddots & t_n & t_{n+1}\\
\vdots &  & \ddots & t_1 & t_0 & t_n\\
t_{n-1} & \cdots & \cdots & t_2 & t_1 & t_0
\end{array}
\right ].}
\label{toeplitz}
\end{equation}%

Eq. (\ref{toeplitz}) is a square matrix of $n\times n$. Supposing that the $i$th, $j$th element of $T$ is denoted by $T_{i,j}$, then the elements in the lower and upper triangular part of matrix $T$ have the relations that $T_{i,j}=T_{i+1,j+1}=t_{i-j}$ and $t_{j-i+n-1}$, $0 \leq i$, $j \leq n-1$. The degrees of freedom of Toeplitz matrix are $2n-1$, rather than $n^2$. Actually, the Toeplitz matrix is not necessarily square. The collision probability of Toeplitz matrix is $n\cdot 2^{-m+1}$, where $n$ is the input length and $m$ is the output length. Toeplitz matrix has low complexity, which supports high speed implementation of privacy amplification.

\subsection{Finite-Size Effect of Privacy Amplification}

The finite-size effect has to be considered to guarantee the security of the final secret keys obtained by CV-QKD systems. We mainly analyse the finite-size effect on the privacy amplification procedure. Assuming that $x$ and $y$ are the classical data of Alice and Bob after they measure their quantum states and $E$ is the quantum states of the eavesdropper Eve. The final secret key rate of a CV-QKD system is expressed as follows~\cite{PE1}:

\begin{equation}
k=\beta I(x:y)-S(y:E)-\Delta(n),
\label{keyrate}
\end{equation}%
where $\beta$ is the reconciliation efficiency, $I(x:y)$ refers to the Shannon entropy of Alice and Bob, $S(y;E)$ refers to the Von Neumann entropy of Bob and Eve for reverse reconciliation, $\Delta(n)$ refers to the effect of finite-size on the security of privacy amplification. Its value can be calculated by~\cite{PE1}:

\begin{equation}
\Delta(n)\equiv (2dim\mathcal{H}_x+3)\sqrt{\frac{log_2(2/\overline{\epsilon})}{n}}+\frac{2}{n}log_2(1/\epsilon_{PA}),
\label{delta}
\end{equation}%
where $dim\mathcal{H}_x$ refers to the dimension of Hilbert space corresponding to the raw key $x$, $\overline{\epsilon}$ is a smoothing parameter, $\epsilon_{PA}$ is the failure probability of privacy amplification, and $n$ is the input block length of privacy amplification. Actually, the parameter $\epsilon_{PA}$ refers to the collision probability of the selected universal hash functions. The parameter $\overline{\epsilon}$ and $\epsilon_{PA}$ are intermediate parameters which can be optimized to a small value that satisfies the security requirement of CV-QKD systems.

As shown in Eq.(\ref{keyrate}), the parameter $\Delta(n)$ has an impact on the final secret key rate. Generally, the secret key rate decreases as the transmission distance increases for CV-QKD systems. Thus, the impact is more obvious on long distance systems. In other words, it will affect the maximum transmission distance of CV-QKD systems. To achieve high secret key rate and support long transmission distance CV-QKD systems, the value of $\Delta(n)$ should be as small as possible. As can be seen from Eq.(\ref{delta}), the input block length has a great influence on the value of $\Delta(n)$ when other factors are fixed. The value of $n$ should be as big as possible so that the $\Delta(n)$ will have a small impact on the final secret key rate. However, if the block length is too long, it will be unrealistic to implement privacy amplification directly. Thus, a compromise block length should be selected. In the previous demonstration, to satisfy the security requirement of privacy amplification when considering the finite-size effect, the block length are respectively required to the order of $10^8$, $10^9$, and $10^{10}$ when the transmission distance are about 50km, 80km, and 100km.

\subsection{The Procedure of Privacy Amplification}

Privacy amplification is used to extract unconditional secure keys from shared weak keys between Alice and Bob and eliminate the information that Eve may know from the side channel. This procedure is performed by choosing a hash function from universal families to compress the shared weak keys and requires a classical authentication channel. As previously described, we choose Toeplitz matrix as hash function to perform the procedure and consider the finite-size effect of privacy amplification. We describe the detailed procedure of privacy amplification as follows.

Assuming that $u_A$ and $u_B$ refer to the shared weak keys for Alice and Bob after error correction with length $n$, $r_A$ and $r_B$ refer to the final secret keys of them with length $l$, and $T$ refers to the Toeplitz matrix of size $n\times l$. The length of strings $u_A$ and $u_B$ is obtained by considering the finite-size effect of privacy amplification. The length of strings $r_A$ and $r_B$ is calculated by $l=\lfloor n\times k \rfloor$, where $k$ is the secret key rate.

{\it Step~1}: Alice randomly generates a uniform string of length $n+l-1$ to construct the Toeplitz matrix $T$. She sends $T$ to Bob. She calculates $r_A=u_AT$ to obtain her final secret keys.

{\it Step~2}: Bob receives $T$ from Alice. He perform the same calculation $r_B=u_BT$ to obtain his final keys.

The strings $u_A$ and $u_B$ are the same after error correction procedure except a small probability that the error correction fails but Alice and Bob can not detect. This probability can be arbitrarily small by performing an error verification procedure. Thus, the final secret keys of Alice and Bob can be considered the same. In practice, an authentication is performed to further ensure the identical of the final secret keys between Alice and Bob, before using them to encrypt or decrypt information in quantum cryptography.

\section{High Speed Implementation of Length-Compatible Privacy Amplification}

High speed privacy amplification is required to support real-time CV-QKD systems. The implementation speed is related to the input length, universal hash functions, implementation platform, calculation precision, and other factors. Because the complexity of Toeplitz matrix is low, we choose it as the hash function. GPU provides parallel computation and it is a suitable platform to perform postprocessing procedure of CV-QKD systems. We have achieved high speed error correction based on it. To simplify the postprocessing implementation procedure, we still implement the privacy amplification process on it. However, the computational complexity is extremely high when performing the long input length privacy amplification directly in software. FFT is used to accelerate the process in our system and it can reduce the complexity from O($n^2$) to O($nlog_2n$). It is difficult and impractical to perform the process directly when the block length is too long. On the other hand, the calculation precision of FFT increases with the block length increases and it affects the correctness and speed of FFT. Thus, length-compatible algorithm is proposed to solve this problem.

\begin{figure}[t]
\centering
\includegraphics[width=36pc]{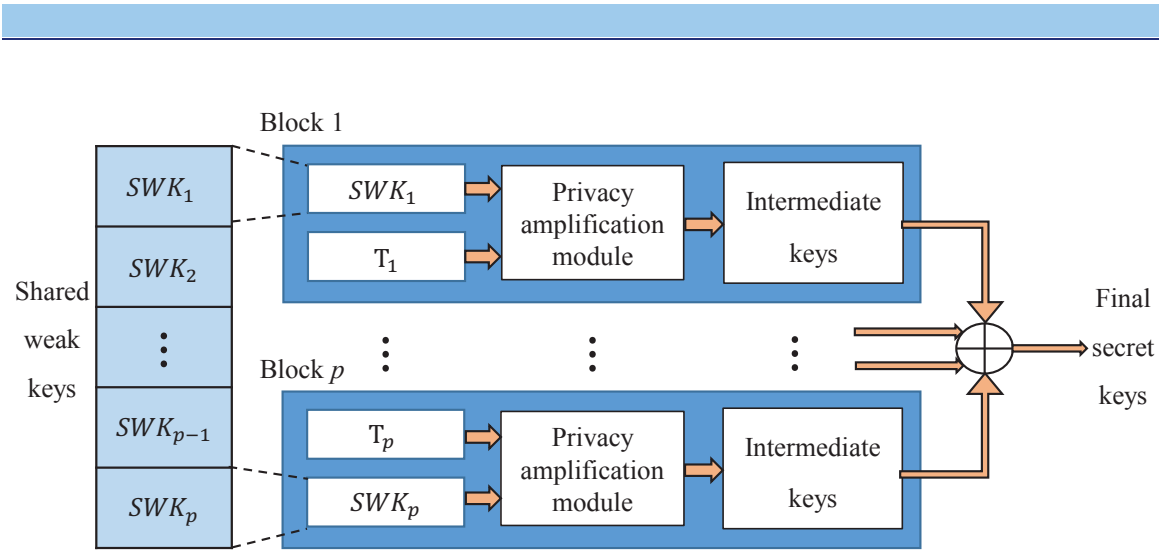}
\caption{The process of length-compatible privacy amplification algorithm. The shared weak keys and the Toeplitz matrix are divided into $p$ blocks. $SWK_i$ and $T_i$ ($i=1,2,\cdots,p$) represent the $i$th block of them. Each of the block individually performs privacy amplification and obtains the intermediate keys. The final secret keys can be obtained by modulo-2 addition among all the intermediate keys.}
\label{PA1}
\end{figure}

The process of length-compatible privacy amplification algorithm is shown in Fig.~\ref{PA1}. The shared weak keys ({\it i.e.} corrected keys) are obtained by the error correction step which has been implemented on GPU. After GPU-based decoding, the corrected keys are copied from device to host. Because we perform the privacy amplification procedure on the GPU platform, the copy process is no longer needed. In a practical CV-QKD system, to achieve unconditional secret keys, the finite-size effect is needed to be considered. The input length of privacy amplification is long, especially in the long distance systems. In practice, the input lengthes are respectively required to the order of $10^8$, $10^9$, and $10^{10}$, when the transmission distance are about 50km, 80km, and 100km. The length will be longer if the distance is further, and it is unrealistic to complete the whole process at one time on a normal platform. Thus, we divide the input data, including the shared weak keys and Toeplitz matrix (universal hash function), into many blocks, then execute them separately to obtain the intermediate keys, finally achieve the final secret keys by modulo-2 addition among all the intermediate keys. The shared weak keys are divided into $p$ blocks sequentially. As shown in Eq.(\ref{Ti}) (The symbol ``$\prime$" represents matrix transpose, and it is the same in Eq.(\ref{Tij})), the Toeplitz matrix is divided into $p$ sub-matrices by rows, and the column of matrix is not divided. If the length of final secret keys is long, the column may also need to be divided.

\begin{equation}
T=[T_0,T_1,\cdots,T_i,\cdots,T_{p-1}]^{\prime}.
\label{Ti}
\end{equation}%

\begin{figure}[t]
\centering
\includegraphics[width=32pc]{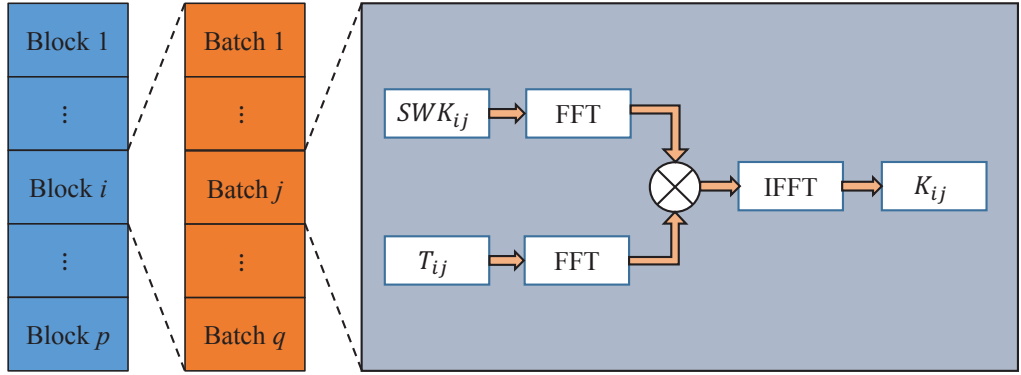}
\caption{The procedure of accelerating length-compatible privacy amplification with FFT. The shared weak keys and the Toeplitz matrix are divided into $p$ blocks. Each block is further divided to $q$ batches. $SWK_{ij}$ and $T_{ij}$ ($i=1,2,\cdots,p$, $j=1,2,\cdots, q$) represent the $j$th batch in the $i$th block. IFFT is the inverse transformation of FFT. $K_{ij}$ refers to the intermediate keys of $j$th batch in the $i$th block. The blocks are serially performed. All the batches in a block are performed in parallel on GPU.}
\label{FFT}
\end{figure}

According to the above method, the privacy amplification of arbitrary length can be realized. Since FFT is used to accelerate the procedure, the calculation precision must match the length. For example, if the length of FFT is lower than $2^{23}$, single-precision floating-point is enough to achieve the correct results. However, if the length is larger than $2^{23}$, double-precision or even long-double-precision must be used, otherwise the results will be wrong. The wrong probability increases with the increase of input length. And the speed of FFT decreases when the precision increases. Thus, to improve the implementation speed and decrease the calculation precision, we further divide each of the blocks ($SWK_i$ and $T_i$, $i=1,2,\ldots,p$) into $q$ batches ($SWK_{ij}$ and $T_{ij}$, $i=1,2,\ldots,p$, $j=1,2,\cdots,q$). The divided method of matrix $T_i$ is shown in Eq.(\ref{Tij}). The computing and memory resources of GPU can not be fully utilized if the length of FFT is too short. To solve this problem, we use multi-batches to independently perform FFT in parallel. The procedure of accelerating length-compatible privacy amplification with FFT is shown in Fig.~\ref{FFT} and is described as follows.

\begin{equation}
T_i=[T_{i0},T_{i1},\cdots,T_{ij},\cdots,T_{i(q-1)}]^{\prime}.
\label{Tij}
\end{equation}%

{\it Step~1}: The length of shared weak keys is extended from $n$ to $2n+l-2$ by filling 0 at the end. Similarly, the length of Toeplitz matrix is extended from $n+l-1$ to $2n+l-2$.

{\it Step~2}: According to the input length of privacy amplification (including the length of extended shared weak keys and extended Toeplitz matrix) and the resources of GPU (mainly refers to the memory size), calculating the number of blocks $p$, then calculating the number of batches $q$.

{\it Step~3}: Performing FFT for each $SWK_{ij}$ and $T_{ij}$, $i=1,2,\ldots,p$, $j=1,2,\cdots,q$, then multiplying the corresponding results of them, next performing IFFT on the results of multiplication, finally the results of IFFT from the $n$th to $(n+k-1)$th are chosen as the intermediate keys $K_{ij}$ of the current batch. The purpose of this step is to calculate all the intermediate keys $K_{ij}$ by using FFT and IFFT. The result of $K_{ij}$ is:

\begin{equation}
{k_{ij}} = SW{K_{ij}} \times {T_{ij}}.
\end{equation}%

The final secret keys are obtained by modulo-2 addition among all the $K_{ij}$.
\begin{equation}
r=\sum_{i=1}^{p}\sum_{j=1}^{q}K_{ij} \quad mod \quad 2.
\label{finalkey}
\end{equation}%

\section{Results}

We implement high speed length-compatible privacy amplification based on GPU, which supports high speed and secure CV-QKD systems and can achieve high secret key rate. The speed is related to the GPU resources, the input length, the calculation precision of FFT, etc. We analyze the effects of these factors in different circumstances. The finite-size effect of privacy amplification is considered to ensure the security of CV-QKD systems. Length-compatible is used to support long input length privacy amplification and decrease the calculation precision of FFT. It is easy to understand that the processing speed of low precision is faster than high precision. Normally, low precision calculation occupies less resources and has a low computational complexity. However, half-precision floating-point number is a new development of nvidia corporation, it needs a special type of GPU to exert its advantages and is not universal to all GPUs. Although our GPU supports half precision calculation, it can not give full play to its advantages. Not only does the processing speed not improve but decline. Thus, we choose single-precision to perform the calculation.

\begin{table}[!t]
\centering
\renewcommand\arraystretch{1.21}
\caption{The speed of privacy amplification under different input length.}
\label{tab1}
\begin{tabular}{|c|ccc|ccc|ccc|}
  \hline
  {\bf Input} & \multicolumn{3}{c|}{\bf Batch size 1Mbits} & \multicolumn{3}{c|}{\bf Batch size 2Mbits} &\multicolumn{3}{c|}{\bf Batch size 4Mbits}\\
  \cline{2-10}
  {\bf length} & {\bf Number of}  & {\bf Time} & {\bf Speed} & {\bf Number of} & {\bf Time} & {\bf Speed} & {\bf Number of} & {\bf Time} & {\bf Speed} \\
  {\bf (Mbits)} & {\bf batches} & {\bf (ms)} & {\bf (Gbps)} & {\bf batches} & {\bf (ms)} & {\bf (Gbps)} & {\bf batches} & {\bf (ms)} & {\bf (Gbps)}\\
  \hline
  4 & 4 & 3.2 & 1.25 & 2 & 3.2 & 1.25 & 1 & 3.2 & 1.25 \\
  8 & 8 & 6.0 & 1.33 & 4 & 6.0 & 1.33 & 2 & 6.0 & 1.33 \\
  16 & 16 & 11.7 & 1.37 & 8 & 11.7 & 1.37 & 4 & 11.7 & 1.37 \\
  32 & 32 & 23.3 & 1.37 & 16 & 23.2 & 1.38 & 8 & 23.2 & 1.38 \\
  64 & 64 & 46.4 & 1.38 & 32 & 46.4 & 1.38 & 16 & 46.5 & 1.38 \\
  128 & 128 & 95.0 & 1.35 & 64 & 95.0 & 1.35 & 32 & 95.1 & 1.35 \\
  \hline
\end{tabular}
\end{table}

Table~\ref{tab1} shows the speed of privacy amplification under different input length. The length of secret keys is 10\% of shared weak keys. In practical systems, the ratio may be a few orders of magnitude lower than 10\%, especially in the case of long distance. As shown in table~\ref{tab1}, the speed has no relationship with the batch size and the number of batches when the input length is consistent. When using single-precision calculation, the results of FFT will be wrong if the input length of a batch is higher than 8Mbits, therefore we set the batch size lower than it. We can achieve the privacy amplification speed to 1.35Gbps at different input length. The results are obtained on a NVIDIA TITAN Xp GPU. However, for practical CV-QKD systems, the finite-size effect of privacy amplification has to be considered to ensure the security. The input length may be over the processing capacity of GPU, we can not perform the procedure directly. Thus, we divide the input data into blocks that can be directly performed. And each block is divided to smaller batches to decrease the calculation complexity and precision. The blocks are processed serially, while the batches are processed in parallel. In fact, if the number of batches is too large or the batch size is too long, the following batches are performed until the previous batches are finished.

\begin{figure}[t]
\centering
\includegraphics[width=32pc]{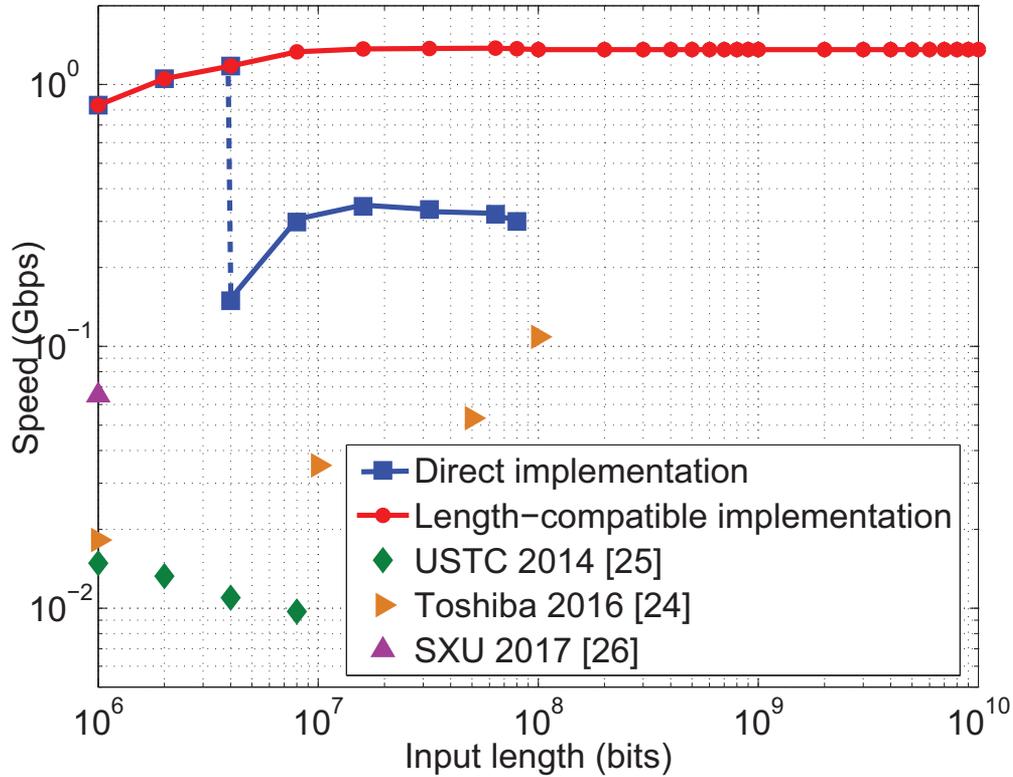}
\caption{The privacy amplification speeds comparison of direct implementation and length-compatible implementation. The red dots represent the speeds of length-compatible implementation. The blue squares represent the speeds of direct implementation. The experiment results between 128Mbits and 10Gbits are obtained on the same batch size with 1Mbits. We still show the previous implementation results of other works. USTC: University of Science and Technology of China. SXU: Shanxi University.}
\label{speed}
\end{figure}

In Fig.~\ref{speed}, we compare the privacy amplification speeds of direct implementation and length-compatible implementation, and we also show the results of other works. The length-compatible implementation can achieve high speed at any input length, where the average speed is about 1.35Gbps. However, only when the single-precision can be used to correctly calculate the results of FFT, direct implementation can achieve high speed. When the input length is between 4M and 80M, direct implementation can obtain the speed to about 0.3Gbps by using double-precision. Due to the limitation of GPU resources, we can not directly perform the privacy amplification procedure when the input length is larger than 80M. In Ref.~\cite{ZhangCM}, they achieve the speeds to about 10Mbps by using multiplication algorithm. In Ref.~\cite{NTT}, they obtain the speed to 108.77Mbps when the input length is 100M by using number theoretic transform to accelerate the privacy amplification procedure, and they implement the procedure on a coprocessor. In Ref.~\cite{FPGA}, they get the speed to 65.443Mbps based on FPGA. As shown in Fig.~\ref{speed}, the proposed length-compatible method can obtain the speed over 1Gbps at any input length, which ensures the security of final secret keys in the case of considering the finite-size effect of privacy amplification and supports high speed real-time CV-QKD system.

\section{Conclusions}
We propose a high speed implementation method of GPU-based length-compatible privacy amplification for continuous-variable quantum key distribution system. The finite-size effect of privacy amplification is considered to ensure secret keys extraction. The long input data is divided into small blocks when GPU can not directly perform the privacy amplification procedure and fast Fourier transform is used to speed up the procedure. To further accelerate the procedure, each block is divided into smaller batches to reduce the calculation precision. The batches are performed in parallel to make full use of the resources of GPU, while the blocks are performed serially. The proposed length-compatible method can be applied to privacy amplification with arbitrary input length. The average speed is achieved to about 1.35Gbps at arbitrary input length, which is one to two orders of magnitude faster than the previous implementation. The early stage of this work has been applied to the longest field test of the continuous-variable quantum key distribution system~\cite{Zhang2017}.

\section*{Acknowledgements}
The authors wish to thank the anonymous reviewers for their valuable suggestions.

\bibliographystyle{IEEEtran}
\bibliography{thesis}

\begin{thebibliography}{10}
\providecommand{\url}[1]{#1}
\csname url@rmstyle\endcsname
\providecommand{\newblock}{\relax}
\providecommand{\bibinfo}[2]{#2}
\providecommand\BIBentrySTDinterwordspacing{\spaceskip=0pt\relax}
\providecommand\BIBentryALTinterwordstretchfactor{4}
\providecommand\BIBentryALTinterwordspacing{\spaceskip=\fontdimen2\font plus
\BIBentryALTinterwordstretchfactor\fontdimen3\font minus
  \fontdimen4\font\relax}
\providecommand\BIBforeignlanguage[2]{{%
\expandafter\ifx\csname l@#1\endcsname\relax
\typeout{** WARNING: IEEEtran.bst: No hyphenation pattern has been}%
\typeout{** loaded for the language `#1'. Using the pattern for}%
\typeout{** the default language instead.}%
\else
\language=\csname l@#1\endcsname
\fi
#2}}

\bibitem{QKD} N.~Gisin, G.~Ribordy, W.~Tittel, and H.~Zbinden, ``Quantum
    cryptography,'' \emph{Rev. Mod. Phys.} vol.~74, no.~1, pp. 145--195,
    2002.

\bibitem{BB84} C. H. Bennett, and G. Brassard, ``Quantum cryptography:
    Public key distribution and cointossing,'' in \emph{Proceedings of IEEE
    Conference on Computer System and Signal Processing,} 1984, pp.
    175-179.

\bibitem{E91} A. K. Ekert, ``Quantum cryptography based on Bell's theorem,''
    \emph{Phys. Rev. Lett.} vol.~67, no.~6, pp. 661--663, 1991.

\bibitem{GG02} F. Grosshans and P. Grangier, ``Continuous variable quantum
    cryptography using coherent states,'' \emph{Phys. Rev. Lett.} vol.~88,
    no.~5, 057902, 2002.

\bibitem{2004NOSW} C. Weedbrook, A. M. Lance, W. P. Bowen, T. Symul, T. C.
    Ralph, and P. K. Lam, ``Quantum cryptography without switching,''
    \emph{Phys. Rev. Lett.} vol.~93, no.~17, 170504, 2004.

\bibitem{2014LZ}Z. Li, Y. C. Zhang, F. Xu, X. Peng, and H. Guo,
    ``Continuous-variable measurement-device-independent quantum key
    distribution,'' \emph{Phys. Rev. A,} vol.~89, no.~5, 052301, 2014.

\bibitem{2015SP}S. Pirandola, C. Ottaviani, G. Spedalieri, C. Weedbrook,
    S. L. Braunstein, S. Lloyd,  T. Gehring, C. S. Jacobsen, and U. L. Andersen, ``High-rate
    measurement-device-independent quantum cryptography,'' \emph{Nat. Photonics}
    vol.~9, pp.397-402, 2015.

\bibitem{2014ZYC}Y. C. Zhang, Z. Li, S. Yu, W. Gu, X. Peng, and H. Guo,
    ``Continuous-variable measurement-device-independent quantum key
    distribution using squeezed states,'' \emph{Phys. Rev. A,} vol.~90, no.~5, 052325, 2014.

\bibitem{2012RMP}C. Weedbrook, S. Pirandola, R. Garcia-Patron, N. J. Cerf,
    T. C. Ralph, J. H. Shapiro, and S. Lloyd, ``Gaussian quantum
    information,'' \emph{Rev. Mod. Phys.} vol.~84, no.~2, pp. 621-669, 2012.

\bibitem{Paul13}P. Jouguet, S. Kunz-Jacques, A. Leverrier, P. Grangier, and
    E. Diamanti, ``Experimental demonstration of long-distance
    continuous-variable quantum key distribution'' \emph{Nat. Photonics}
    vol.~7, pp.378-381, 2013.

\bibitem{2015en}E. Diamanti, and A. Leverrier, ``Distributing secret keys
    with quantum continuous variables: principle, security and
    implementations,'' \emph{Entropy}, vol.~17, no.~9, pp. 6072-6092, 2015.

\bibitem{Zhang2017}Y. C. Zhang, Z. Li, Z. Chen, C. Weedbrook, Y. Zhao, X.
    Wang, C. Xu, X. Zhang, Z. Wang, M. Li, X. Zhang, Z. Zheng, B. Chu, X.
    Gao, N. Meng, W. Cai, Z. Wang, G. Wang, S. Yu, and H. Guo,
    ``Continuous-variable QKD over 50km commercial fiber,'' Preprint at arXiv:1709.04618, 2017.

\bibitem{PE1} A. Leverrier, F. Grosshans and P. Grangier, ``
    Finite-size analysis of a continuous-variable quantum key distribution,''
    \emph{Phys. Rev. A,} vol.~81, no.~6, 062343, 2010.

\bibitem{PE2} P. Jouguet, S. Kunz-Jacques, E. Diamanti and A. Leverrier,
    ``Analysis of imperfections in practical continuous-variable
    quantum key distribution,'' \emph{Phys. Rev. A,} vol.~86, no.~3, 032309, 2012.

\bibitem{IR1} G. V. Assche, J. Cardinal and N. J. Cerf, ``
    Reconciliation of a quantum-distributed Gaussian key,'' \emph{IEEE Transactions
     on Information Theory,} vol.~50, no.~2, pp. 394-400, 2004.

\bibitem{WangRA}X. Wang, Y. C. Zhang, Z. Li, B. Xu, S. Yu, and H. Guo,
    ¡°Efficient rate-adaptive reconciliation for continuous-variable quantum
    key distribution,¡± \emph{Quantum Inf. Comput.} vol.~17, no.~13\&14, pp. 1123-1134, 2017.

\bibitem{WangHSEC}X. Wang, Y. C. Zhang, S. Yu, and H. Guo,
    ``High speed error correction for continuous-variable quantum key distribution with
    multi-edge type LDPC code,'' Preprint at arXiv:1711.01783, 2017.

\bibitem{PA1}C. H. Bennett, G. Brassard, and J. M. Robert,
    ``Privacy amplification by public discussion,'' \emph{SIAM journal on
    Computing,} vol.~17, no.~2, pp. 210-229, 1988.

\bibitem{PA2} C. H. Bennett, G. Brassard, C. Cr\'{e}peau and U. M. Maurer,
    ``Generalized privacy amplification,'' \emph{IEEE
    Transactions on Information Theory,} vol.~41, no.~6, pp. 1915-1923, 1995.

\bibitem{PA3} D. Deutsch, A. Ekert, R. Jozsa, C. Macchiavello, S. Popescu
    and A. Sanpera, ``Quantum privacy amplification and the
    security of quantum cryptography over noisy channels,'' \emph{Phys. Rev.
    Lett.} vol.~77, no.~13, pp. 2818-2821, 1996.

\bibitem{Hash1}M. N. Wegman, and J. L. Carter, ``New hash functions
    and their use in authentication and set equality,'' \emph{J. Comput.
    Syst.} vol.~22, no.~3, pp. 265-279, 1981.

\bibitem{Hash2}D. R. Stinson, ``Universal hash families and the
    leftover hash lemma, and applications to cryptography and computing,''
    \emph{Journal of Combinatorial Mathematics and Combinatorial Computing,} vol.~42,
    pp. 3-32, 2002.

\bibitem{Toeplitz}H. Krawczy, ``Lfsr-based hashing and authentication,''
    in \emph{Proceedings of Advances in Cryptology - CRYPTO94, International Cryptology
    Conference,} 1994, pp. 129-139.

\bibitem{NTT}R. Takahashi, Y. Tanizawa, and A. R. Dixon,
    ``High-speed implementation of privacy amplification in quantum key
    distribution,'' In \emph{6th Int. Conf. Quantum Cryptography,}
     Washington, DC, USA, 2016,  presentation number: 160.

\bibitem{ZhangCM}C. M. Zhang \emph{et al.}, ``Fast implementation
    of length-adaptive privacy amplification in quantum key distribution,''
    \emph{Chinese Physics B,} vol.~23, no.~9, 090310, 2014.

\bibitem{FPGA}S. S. Yang, Z. L. Bai, X. Y. Wang, and Y. M. Li, ``FPGA-Based
    Implementation of Size-Adaptive Privacy Amplification
    in Quantum Key Distribution,'' \emph{IEEE Photonics Journal,} vol.~9, no.~6, pp. 1-8, 2017.

\bibitem{long11}P. Jouguet, S. Kunz-Jacques, and A. Leverrier,
    ``Long-distance continuous-variable quantum key distribution with a
    Gaussian modulation,'' \emph{Phys. Rev. A,} vol.~84, no.~6, 062317, 2011.

\bibitem{Slice1}J. Lodewyck, M. Bloch, R. Garc\'{\i}a-Patr\'{o}n, S.
    Fossier, E. Karpov, E. Diamanti, T. Debuisschert, N. J. Cerf, R.
    Tualle-Brouri, S. W. McLaughlin, and P. Grangier, ``Quantum key
    distribution over 25 km with an all-fiber continuous-variable system,''
    \emph{Phys. Rev. A,} vol.~76, no.~4, 042305, 2007.

\bibitem{MR}A. Leverrier, R. All\'{e}aume, J. Boutros, G. Z\'{e}mor, and P.
    Grangier, ``Multidimensional reconciliation for a continuous-variable
    quantum key distribution,'' \emph{Phys. Rev. A,} vol. 77, no.~4, 042325
    2008.

\bibitem{METLDPC}T. Richardson and R. Urbanke, ``Multi-edge type LDPC
    codes,'' presented at \emph{Workshop honoring Prof. Bob McEliece on his 60th
    birthday,} California Institute of Technology, Pasadena, California, 2002, pp.
    24-25.

\bibitem{ZhangXY}X. Zhang, Y. Zhang, Y. Zhao, X. Wang, S. Yu, and
    H. Guo, ``Finite-size analysis of continuous-variable
    measurement-device-independent quantum key distribution,'' \emph{Physical Review
    A,} vol.~96, no.~4, 042334, 2017.

\end{thebibliography}

\end{document}